\documentclass[aps,prb,twocolumn,multicol,superscriptaddress,floats,showpacs]{revtex4-1}
\usepackage{bm}

\usepackage{amssymb}
\usepackage{pifont}
\usepackage{amsmath}
\usepackage{graphics}
\usepackage{graphicx}
\usepackage{epsfig}
\usepackage{pstricks}
\usepackage{pst-plot}
\usepackage{pst-slpe}
\usepackage{latexsym}
\usepackage[hang, nooneline]{subfigure}

\newcommand{\be}{\begin{equation}}
\newcommand{\ee}{\end{equation}}
\newcommand{\bea}{\begin{eqnarray}}
\newcommand{\eea}{\end{eqnarray}}
\newcommand{\phrl}[1]{Phys.~Rev.~Lett. {\bf #1}}
\newcommand{\phrb}[1]{Phys.~Rev.~B {\bf #1}}

\newcommand{\q}{{\bf q}}
\renewcommand{\k}{{\bf k}}

\begin{document}
\title{Inhomogeneous longitudinal electric field-induced anomalous Hall conductivity in a ferromagnetic two-dimensional electron gas}
\author{Ankur Sensharma}
\affiliation{Department of Theoretical Physics, Indian Association for the Cultivation of Science,Jadavpur, Kolkata 700 032, India}
\affiliation{Department of Physics, University of Gour Banga, Malda 732101, West Bengal, India}
\author{Sudhansu S. Mandal}
\affiliation{Department of Theoretical Physics, Indian Association for the Cultivation of Science,Jadavpur, Kolkata 700 032, India}
%


\date{\today}

\begin{abstract}
It is known that the anomalous Hall conductivity (AHC) in a disordered two dimensional electron system with Rashba spin-orbit interaction
and finite ferromagnetic spin-exchange energy is zero in the metallic weak-scattering regime because of the exact cancellation of 
the bare-bubble contribution
by the vertex correction. We study the effect of inhomogeneous longitudinal electric field on the AHC in such a system. We predict that
AHC increases from zero (at zero wavenumber), forms a peak, and then decreases as the wavenumber for the variation of electric
field increases. The peak-value
of AHC is as high as the bare-buble contribution. We find that the wave number, $q$, at which the peaks occur is the inverse
of the geometric mean of the mean free path of an electron and the spin-exchange length scale. Although the Rashba energy is responsible for the
peak-value of AHC, the peak position is independent of it.

\end{abstract}
\maketitle
\section{Introduction}


Hall resistance in a ferromagnetic material is known to have two contributions: 1) proportional to magnetic field,
2) proportional to magnetization. While the former is responsible for conventional Hall effect, the latter is due to
anomalous Hall effect (AHE). The anomalous Hall coefficient is the zero field Hall resistance which is estimated by
extrapolating Hall resistance up to zero magnetic field. The physical mechanisms which contribute to the anomalous
Hall conductivity (AHC), are broadly
categorized into two classes: intrinsic and extrinsic. While the properties of the bands are responsible for the
former, the spin dependent scattering of electrons belong to the extrinsic mechanism. Karplaus and Luttinger \cite{KL54} proposed
intrinsic mechanism due to the spin-orbit interaction (SOI) for the polarized electrons in ferromagnetic materials. This
mechanism was later realized as the outcome of the Berry phase \cite{Berry84} from band topology. Smit \cite{Smit55} proposed
an extrinsic mechanism which is known as `skew-scattering' (SS) in which electrons scatter from impurities with spin asymmetry.
In addition, contribution due to side-jump (SJ) scattering was put forward by Berger~\cite{Berger70} through a semiclassical theory.


The debate on the true mechanism for AHE grew stronger with the contradicting results \cite{Review10} for various systems. The linear 
transport theories, which are now mostly used to study AHE, seem to have settled some of the confusing issues in the recent
years. Although, different approaches of this theory often tend to conflict with each other, there are some successful
attempts \cite{Sinitsyn07,KovalevRapid} to establish links between them. To this end, the two dimensional ferromagnetic-Rashba gas,
{\it i.e.}, electrons with Rashba SOI and ferromagnetic exchange coupling 
has proved to be a simple and elegant model to work with because it captures almost all the important and fundamental features
relevant to AHE.  Due to the Rashba SOI, characterized by the Rashba energy $\Delta_{_R}$, degeneracy is lifted by the formation of 
two chiral subbands with opposite helicity. Further,
ferromagnetic spin-exchange energy produces a gap between these subbands.

A number of studies on AHC have been performed within the model of two-dimensional ferromagnetic-Rashba gas, using the Kubo formalism
~\cite{Bruno01,Dugaev05,Inoue06,Sinova07,Sinitsyn07,Borunda07,Kato07,Zhou10},the Keldysh technique
~\cite{Onoda06,Sinitsyn07,Kovalev09}, and the semiclassical treatment of Boltzmann transport equation
~\cite{Culcer03,Liu06}. According to the studies in Kubo formalism, the magnitude of AHC is sensitive to whether only one or both
the subbands are occupied by the electrons, {\it i.e.}, whether or not the Fermi level is situated
 below the band-gap created by exchange energy, $\Delta_{ex}$.
In the presence of non-magnetic impurities, both the Fermi sea and Fermi surface contributions in AHC vanish \cite{Sinova07}
when $\epsilon_{_F}>\Delta_{ex}$,
where $\epsilon_{_F}$ is the Fermi energy. The bare-buble contribution is exactly canceled by the vertex correction
\cite{Inoue06,Sinova07} due to SJ mechanism, and the SS term is absent \cite{Borunda07}. 
 On the contrary, all these terms are finite for $\epsilon_{_F}<\Delta_{ex}$. A comprehensive analysis of these works
can be found in the detailed study of Nunner {\it{et al}}~\cite{Sinova07}.
Using the Keldysh technique in a T-matrix approximation, Onoda {\it et al}\cite{Onoda06,Onoda08}, however,  find a strong SS contribution that
leads to non-zero AHC even when  $\epsilon_{_F}>\Delta_{ex}$.
They also observe that the AHC changes sign when the Fermi
level enters into the band gap. The Keldysh technique automatically incorporates the higher order Born scatterings and enables
them to arrive at such striking conclusions when the Fermi energy is located around the anticrossing of band dispersions in
momentum space. Kovalev {\it{et al}}~\cite{Kovalev09} later included the higher order scatterings explicitly in the Kubo formalism
to match the results of Onoda {\it{et al}}~\cite{Onoda06} and were able to unify the these approaches~\cite{KovalevRapid}
at least for a certain range of parameters.

There is no controversy, however, in the metallic weak-scattering regime $(\epsilon_{_F}\tau >1)$, where $\tau$ is the mean scattering time
of an electron.   When the Fermi
level goes far above the band gap ($\epsilon_{_F}>\Delta_{ex})$, {\it i.e.}, when both the chiral subbands are occupied,
the AHC vanishes~\cite{Review10,Sinova07,InoueR09,KovalevRapid},
making this regime a somewhat uninteresting one. We however note that AHE is intimately related to
spin hall effect \cite{Lee04,Schwab10} and there are instances of enhancement of 
spin-Hall conductivity \cite{Mandal08} and occurrence of a induced transverse force~\cite{Ballentine77} due to
 inhomogeneity of the applied electric field. 
In this paper, we study the effect of inhomogeneous longitudinal electric field
($\bm{E}_{\bm{q}} \parallel \bm{q}$) on the AHC in the metallic weak-scattering regime when both
the subbands are occupied.

We find, within the Kubo formalism, that AHC, $\sigma^{AH}$, is zero at $q=0$, increases with $q$ and forms a peak at certain $q$ before it decreases to zero at large $q$.
$\sigma^{AH}$ depends on the parameters $\epsilon_{_F}\tau$, $\Delta_{ex}\tau$, and $\Delta_R\tau$.
We find that the peak-value of AHC is proportional to $(\Delta_R\tau)^2(\epsilon_{_F}\tau)^{-1}$ and the peak occurs when
$ql_{eff} \approx 1$, where $l_{eff} = (ll_{ex})^{1/2}$, $l$ is the mean free path of an electron, and $l_{ex} = v_{_F}/(4\Delta_{ex})$
is the length-scale corresponding to spin-exchange, with $v_{_F}$ being the Fermi velocity.
We note that the momentum at which the peaks occur does not depend on $\Delta_{_R}$.



The paper is organized as follows. The following section contains a brief formalism for the evaluation
of the anomalous Hall conductivity. In Section III, we present our results and discuss about the relevant scales and limits.
Finally, we summarize our results in Section IV.

\section{Anomalous Hall Conductivity}
\subsection{The model}
A system of spin-polarized two dimensional electron gas with Rashba spin-orbit interaction in a disordered environment 
may be expressed by the Hamiltonian
\begin{equation}
H =\left(\frac{\vec{\nabla}^2}{2m}+V(r)\right) \sigma_0 + \alpha (\vec{\sigma} \times \vec{\nabla})\cdot \hat{z}
 - \Delta_{ex} \sigma_z  \, ,
\end{equation}
where $m$ is the the effective mass of an electron, $\alpha$ is the Rashba spin-orbit coupling parameter, $\Delta_{ex}$ 
is the exchange energy which favors one kind of spin over the other, $\vec{\sigma}$ represents three Pauli matrices 
and $\sigma_0$ is $2\times 2$ unit matrix. Here $V(\bf r)$ is the spin independent disorder potential
for the randomly located $\delta$-function impurities and has the form $V({\bf r}) =\sum_i V_i \delta ({\bf
  r}-{\bf R}_i)$ satisfying $\langle V_i \rangle =0$ , $\langle V_i^2\rangle =V_2 \ne 0$ and 
$\langle V_i^3\rangle =V_3 \ne 0$. (We have set the unit $\hbar =1$ and $c=1$.)

In the absence of disorder, the above Hamiltonian can be exactly solved with the eigen values
\begin{equation}
E_{k}^s = \frac{k^2}{2m} + s\zeta_k  
\end{equation}
where $k$ is the momentum of an electron, $s= \pm$ is the label for two chiral subbands produced due to Rashba spin-orbit
interaction and $\zeta_k=\sqrt{\Delta_{ex}^2 + \alpha^2 k^2}$. The corresponding Fermi momenta $K_s$ are given by
\begin{equation}
 K_s^2=2m\left[ \epsilon_{_F}-s\zeta_s\right]
 \label{eq:fermi_momenta}\,
\end{equation}
where $\epsilon_{_F}$ is the Fermi energy and 
\begin{equation}
 \zeta_s = \sqrt{\Delta_{ex}^2+\Delta_R^2+m^2\alpha^4} - s \, m\alpha^2
\, ,
\end{equation}
with $\Delta_R=\alpha\sqrt{2m\epsilon_{_F}}=\alpha k_{_F}$. Here $k_{_F}$ is the mean Fermi momentum,
given by $\epsilon_{_F}=\frac{k_{_F}^2}{2m}$.
The density of states at the Fermi level for the two subbannds are 
\begin{equation}
\nu_s=\nu_0
\left[ 1-s\frac{m\alpha^2}{\sqrt{\Delta_{ex}^2+\Delta_R^2+m^2\alpha^4}}\right] \, .
\end{equation}
where $\nu_0$ is the density of states for each spin in pure two dimensional electron gas.
The free retarded (advanced) Green's function of the system can then be expressed as
\begin{equation}
\hat{G}_{\k ,0}^{R,A}(\epsilon) = \frac{1}{2}
\sum_{s=\pm} \frac{\sigma_0 + s \alpha  (\bm{\sigma}\times \k)\cdot \hat{z} 
- \Delta_{ex}\sigma_z)/\zeta_k}{\epsilon - \xi_\k^s \pm i\eta} \, , 
\end{equation}
where $\xi_\k^s=E_\k^s-\epsilon_F$.
The self energy for the electrons due to scattering is then found to be
\begin{equation}
\Sigma^{R(A)}=\mp\frac{i}{4\tau \nu_0}\left[(\nu_{+}+\nu_{-})\sigma_0-\Delta_{ex}\left(\frac{\nu_+}{\zeta_+}-\frac{\nu_-}{\zeta_-}\right)\sigma_z\right]
\label{selfenergy}  
\end{equation}
in Born approximation, 
where $\tau = (2\pi\nu_0n_iV_2)^{-1}$ is the mean scattering time for an electron, and $n_i$ represents the concentration of impurity.
We thus find the disorder-averaged Greens functions as
\begin{equation}
\hat{G}_{\k}^{R}(\epsilon) = \frac{1}{2}
\sum_{s=\pm} \frac{\sigma_0 + s \alpha  (\bm{\sigma}\times \k)\cdot \hat{z} 
- \tilde{\Delta}_{ex}\sigma_z)/\tilde\zeta_k}{\tilde{\epsilon} - \tilde\xi_\k^s } \, , 
\end{equation}
where $\tilde{\epsilon} = \epsilon + i (\nu_+ +\nu_-)/(4\tau\nu_0)$ and 
$\tilde{\Delta}_{ex} = \Delta_{ex} \left[1 +\frac{i}{4\tau \nu_0} \left(  \frac{\nu_+}{\zeta_+}-\frac{\nu_-}{\zeta_-} \right) \right] $.
Here $\xi_\k^s$ and $\zeta_k$ get renormalized to $\tilde\xi_\k^s$ and $\tilde\zeta_k$ due to the presence of $\tilde\Delta_{ex}$
instead of $\Delta_{ex}$ in their respective expressions.
When both the helicity subbands are occupied, $\nu_+ +\nu_- = 2\nu_0$ and $\nu_+/\zeta_+ = \nu_-/\zeta_-$ and therefore exchange energy
does not get renormalized and $\tilde{\epsilon}$ becomes $\epsilon + i/2\tau$.

\subsection{Kubo Formula}
\begin{figure}
\begin{minipage}{\columnwidth}
\vspace*{0.5cm}
\includegraphics[scale=.6]{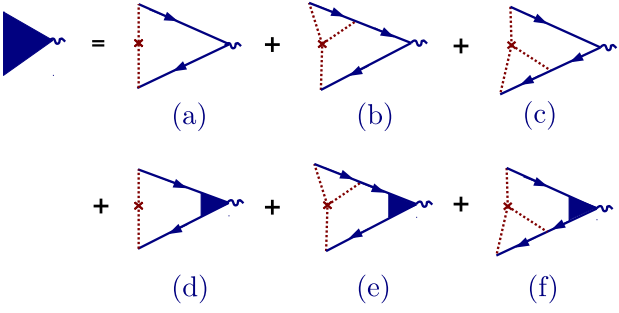}
\end{minipage}
\caption{(Color online) Diagrammatic representation of the self-consistent equation [\ref{corrected vertex}] including side-jump (ladder)
and skew-scattering corrections.
The small wavy lines with a point vertex represent the bare charge current vertex $\hat j_x$,
and the small wave lines with dressed vertex represent $\hat\Gamma_x$. Retarded and advanced Green's functions are represented
by the solid lines with arrows in the opposite directions. Dashed lines refer to the impurity scattering. 
Lowest order  side-jump (a)
and skew-scattering (b,c) diagrams; 
diagrams (d,e,f) represent vertex correction to the diagrams (a,b,c) respectively. Here vertex correction 
due to both side-jump and skew-scattering.}
\label{fig:fullvertex}
\end{figure}

Using Kubo formula, anomalous Hall conductivity may be expressed as
\begin{eqnarray}
\sigma^{AH}_{yx}\!\!&=& \!\!\frac{1}{2\pi}\int \frac{d\k}{(2\pi)^2}\,{\rm Tr}\,\left[\hat j_y\left({\k}+\frac{\q}{2}\right) 
\hat G^R_{\k+\q}(0)\right. \nonumber\\
& \times &\left. \left\{\hat j_x\left({\k}+\frac{\q}{2}\right)+\hat \Gamma_x({\q})\right\}\hat G^A_{\k}(0)\right]\,  
\label{sigmatotal}
\end{eqnarray}
where charge current operators $\hat j_x({\k})=e(\frac{k_x}{m}\sigma_0-\alpha\sigma_y)$ and 
$\hat j_y({\k})=e(\frac{k_y}{m}\sigma_0+\alpha\sigma_x)$, and $\hat \Gamma_x({\q})$ represents the correction to the vertex of $\hat j_x$
due to both side-jump
and skew scattering contributions (see fig.\ref{fig:fullvertex}) which can be calculated by solving
the self-consistent equation

\begin{eqnarray}
\hat\Gamma_x({\q}) &=& \gamma_1\int\frac{d\k'}{(2\pi)^2}\hat G^R_{\k'+\q}(0)
\left[ \hat j_x\left({\k'}+\frac{\q}{2}\right)+\hat\Gamma_x({\q})\right] \hat G^A_{\k'}(0)  \nonumber\\
&+&\gamma_2 \int\frac{d\k'}{(2\pi)^2}\hat G^R_{\k'+\q} \int\frac{d\k''}{(2\pi)^2}\hat G^R_{\k''+\q} \nonumber \\
& & \,\,\,\,\,\, \,\,\,\times \left[ \hat j_x\left({\k''}+\frac{\q}{2}\right)+\hat\Gamma_x(\q)\right] \hat G^A_{\k''}(0) \nonumber\\
&+&\gamma_2 \int\frac{d\k''}{(2\pi)^2}\hat G^R_{\k''+\q}(0) 
 \left[ \hat j_x\left({\k''}+\frac{\q}{2}\right) 
      +\hat\Gamma_x(\q)\right] \nonumber \\
& & \,\,\,\,\,\,\,\, \times \, \hat G^A_{\k''}(0)
\int\frac{d\k'}{(2\pi)^2}\hat G^A_{\k'}(0) ,
\label{corrected vertex}
\end{eqnarray}
where, $\gamma_1=n_iV_2$ and $\gamma_2=n_iV_3$.
It is established from a number of studies ~\cite{Sinitsyn07,Borunda07,Sinova07} that the skew-scattering contribution to
the homogeneous $(q=0)$ anomalous Hall conductivity vanishes provided both the chiral subbands are occupied. This, however, is true for any $q$
since the last two terms in Eq.(\ref{corrected vertex}) cancel each other because of the identity
 $\int\frac{d\k}{(2\pi)^2}\hat G^{R,A}_{\k}=\mp i\pi\nu_{_0}\sigma_{_0}$. 
Equation (\ref{sigmatotal}) therefore reduces to
\begin{eqnarray}
\sigma^{AH}_{yx}\!\!&=& \!\!\frac{1}{2\pi}\int \frac{d\k}{(2\pi)^2}\,{\rm Tr}\, \left[\hat j_y\left({\k}+\frac{\q}{2}\right) 
\hat G^R_{\k+\q}(0)\right. \nonumber\\
& &\times \,\left.  \left\{\hat j_x\left({\k}+\frac{\q}{2}\right)+\hat J_x({\q})\right\}\hat G^A_{\k}(0)\right]
\label{sigmaladder}
\end{eqnarray}
with $\hat J_x({\q})$ containing side-jump types of scattering contribution of $\hat\Gamma_x({\q})$ only, {\it i.e.},
\begin{eqnarray}
\hat J_x({\q})&=&\frac{1}{2\pi\nu_{_0}\tau}\int\frac{d\k'}{(2\pi)^2}\hat G^R_{\k'+\q} (0)
\left\{\hat j_x\left({\k}+\frac{\q}{2}\right) \right. \nonumber \\
 & & \left. +\hat J_x({\q})\right\}\hat G^A_{\k'} (0)\, .
\label{Jv}
\end{eqnarray}
We numerically evaluate $\hat J_x (\q)$ and hence $\sigma_{yx}^{AH}$ below.

\subsection{Numerical evaluation}
The numerical evaluation of $\sigma^{AH}_{yx}$ {\it via} Eqs. (\ref{sigmaladder}) and (\ref{Jv}) proceeds with the same spirit
of Refs.~\onlinecite{Mandal08,Zhou10}. Expanding $\hat J_x({\q})$ in the Pauli matrix basis:
 $\hat J_x=\sum_{\alpha =0}^3 J_{\alpha}\sigma_{\alpha}$, Eq.(\ref{Jv}) can be rewritten as
\begin{eqnarray}
& & \sum_{\alpha} J_{\alpha}({\q})\sigma_{\alpha}
-\frac{1}{2\pi\nu_{_0}\tau}\int\frac{d\k}{(2\pi)^2}\hat G^R_{\k+\q}(0)\left(\sum_{\alpha} J_{\alpha}({\q})\sigma_{\alpha}\right) \nonumber \\
&& \times \hat G^A_{\k}(0) 
=\frac{1}{2\pi\nu_{_0}\tau}\int\frac{d\k}{(2\pi)^2}\hat G^R_{\k+\q}(0)\hat j_x({\k}+\frac{\q}{2})\hat G^A_{\k} (0).
\label{rnkgmk}
\end{eqnarray}
Equating the coefficients of the four Pauli matrices in the above equation, we find a $4\times 4$ matrix equation:
\begin{equation}
\left(
\begin{array}{cccc} 
1-m_{00}& -m_{11} & -m_{22} & -m_{33} \\
-m_{01}& 1-m_{10} & im_{23} & -im_{32} \\ 
-m_{02}& -im_{13} & 1-m_{20} & im_{31} \\
-m_{03}& im_{12} & -im_{21} & 1-m_{30}
\end{array}
\right)
\left(
\begin{array}{c}
 J_0\\ J_1\\ J_2\\ J_3
\end{array}
\right)
=
\left(
\begin{array}{c}
  z_0\\ z_1\\ z_2\\ z_3
 \end{array}
\right) \, ,
\label{matrix}
\end{equation}
where $m_{\alpha\beta} (\q)$ and $z_{\beta}(\q)$ are given by \cite{Sensharma06}
\begin{equation}
\sum_{\beta=0}^3 m_{\alpha\beta}\sigma_{\beta}=
\frac{1}{2\pi\nu_{_0}\tau}\int\frac{d\k}{(2\pi)^2}\hat G^R_{\k+\q}\sigma_{\alpha}\hat G^A_{\k}\sigma_{\alpha}
\label{matrix1}
\end{equation}
\begin{equation}
z_\beta(\q)=\frac{1}{4\pi\nu_{_0}\tau}\,{\rm Tr}\,
\left[\int\frac{d\k}{(2\pi)^2}\hat G^R_{\k+\q}(0)\hat j_x\left({\k}+\frac{\q}{2}\right)\hat G^A_{\k}(0)
\sigma_\beta\right].
\label{matrix2}
\end{equation}
Inverting the matrix equation (\ref{matrix}), we find $J_\alpha$ and then substituting these in Eq.(\ref{sigmaladder}),
we obtain $\sigma_{yx}^{AH}$. 


\section{results and discussions}

 
In the metallic regime ($\epsilon_{_F}\tau\gg 1$), there is no net $\sigma_{yx}^{AH}$ when the applied electric field is
uniform. When both subbands are partially occupied, the bare bubble contribution
\be
\sigma^{AH}_b=-\frac{e^2}{2\pi}\frac{m\alpha^2\Delta_{ex}}{\Delta_{ex}^2+\Delta_R^2+(\frac{1}{2\tau})^2}
\label{sigmabare}
\ee
is exactly canceled by the ladder vertex correction,$\sigma^{AH}_v$, which is numerically equal to $\sigma^{AH}_b$ but
opposite in sign. In the super clean limit ($\tau\rightarrow\infty$), the above value agrees with the previous result\cite{Sinova07}.

\begin{figure}[h]
\centering
\includegraphics[scale=0.5]{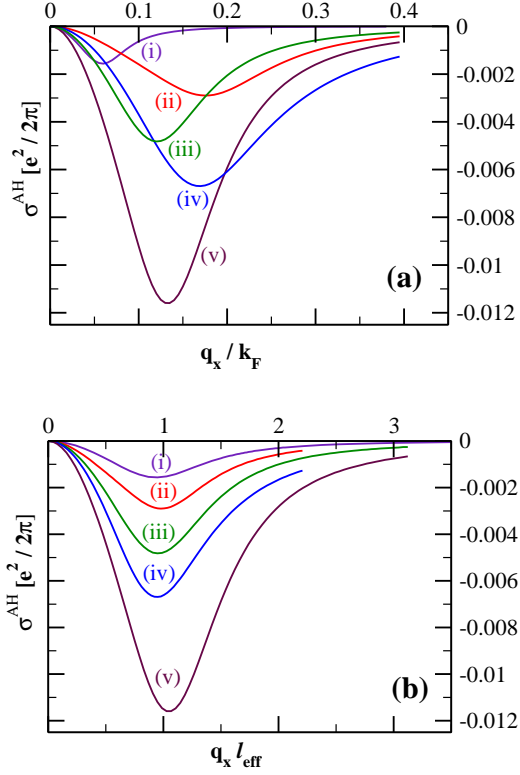}
\caption{(Color online) (a) The variation of anomalous Hall conductivity in the unit of $e^2/(2\pi)$ with $q_x/k_{_F}$ for different sets
of parameters (i)--(v) shown in Table-1. The peaks in $\sigma^{AH}$ occur at different valued of $q_x/k_{_F}$ depending on the choice of
the parameters. Scaling $q_x/k_{_F}$ by $(\epsilon_{_F}\tau)^{-1}(\Delta_{ex}\tau)^{-1/2}$, $\sigma^{AH}$ is replotted (b) against $q_xl_{_{eff}}$
for all the chosen sets of parameters. Peaks in (b) occur when $q_xl_{_{eff}} \sim 1$.}
\label{AHE-line}
\end{figure}

\begin{table}[t]
\newcommand{\mc}[3]{\multicolumn{#1}{#2}{#3}}
\begin{tabular}{|c|c|c|c|c|}\cline{1-5}
 Set & $\Delta_{_R}\tau$ & $\Delta_{ex}\tau$ & $\epsilon_{_F}\tau$ & $-\sigma^{AH}_b [e^2/(2\pi)]$\\\hline\hline
\mc{1}{|l|}{(i)}&0.2 & 0.4 & 10.0 & 0.0018\\\hline
\mc{1}{|l|}{(ii)}&0.2 & 0.8 & 5.0 & 0.0034\\\hline
\mc{1}{|l|}{(iii)}&0.25 & 0.4 & 5.0 & 0.0061\\\hline
\mc{1}{|l|}{(iv)}&0.2 & 0.2 & 2.5 & 0.0071\\\hline
\mc{1}{|l|}{(v)}&0.4 & 0.4 & 5.0 & 0.0112\\\hline

\end{tabular}
\label{parm} 
\caption{Five sets (i)--(v) of three parameters $\Delta{_R}\tau$, $\Delta_{ex}\tau$, and $\epsilon_{_F}\tau$ used to calculate 
$\sigma^{AH}$ for the Fig.~(\ref{AHE-line}). The last column shows the value of $\sigma^{AH}_b$
calculated using Eq.(\ref{sigmabare}) in the unit of $e^2/(2\pi)$ for these sets of parameters.}
\end{table}

\begin{figure}[h]
\centering
\includegraphics[scale=0.4]{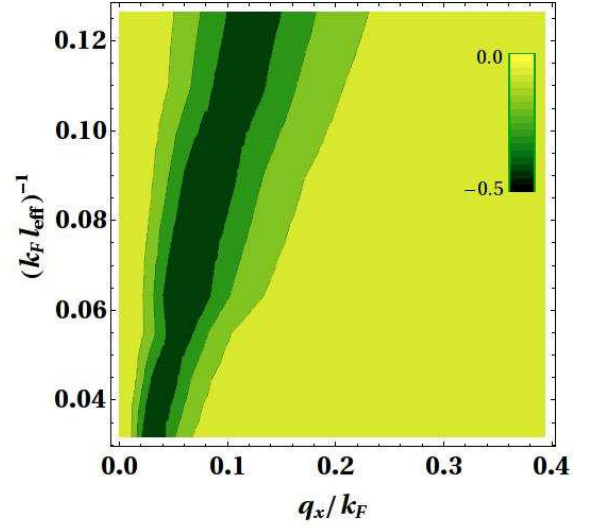}
\caption{(Color online) Contour plot of scaled anomalous Hall conductivity,
$\tilde\sigma^{AH}=\sigma^{AH}(\epsilon_{_F}\tau)(\Delta_R\tau)^{-2}$ versus
$\frac{q_x}{k_F}$ and $(k_Fl_{_{eff}})^{-1}$.  The darkest linear patch, where the magnitude of $\tilde\sigma^{AH}$ is maximum, 
occurs when $q_xl_{_{eff}}\approx 1$.}  
\label{AHE-cont}
\end{figure}

In this paper, we are considering the effect of inhomogeneous longitudinal electric field, {\it i.e.,} $\q \parallel \bm{E}_\q$ 
on the Anomalous Hall conductivity. We define three dimensionless parameters out of four energy scales $\epsilon_{_F}$,
$\Delta_R$, $\Delta_{ex}$ and $1/\tau$. They are $\epsilon_{_F}\tau,\Delta_R\tau$ and $\Delta_{ex}\tau$.
We numerically calculate $\sigma_{yx}^{AH}$ for different sets of these parameters given in Table-1, 
using Eqs.~(\ref{sigmaladder}), (\ref{matrix})--(\ref{matrix2}) when the electric field is assumed to be applied along x-axis, i.e., $E_x = E_0e^{iq_xx}, E_y =0$. 

As evident in Fig.\ref{AHE-line}(a), the
magnitude of $\sigma^{AH}$ increases, forms a peak, and then decreases with the increase of $q _x$. We note that
the peak values for the different sets of parameters are close to the corresponding bare contributions $\sigma^{AH}_b$ listed in Table-1.
This suggests that the effect of inhomogeneous electric field on the AHC is substantial.
The peak positions and the maximum values of $\sigma^{AH}$ depend on these three parameters. However, if we scale $q_x$ in the unit of inverse
effective mean free path, $l_{eff} = \sqrt{l l_{ex}}$ where $l =v_{_F}\tau$ is the mean free path of an electron and 
$l_{ex}= v_{_F}/(4\Delta_{ex})$ is
the length scale corresponding to the exchange energy, 
the peak positions occur at $q_xl_{eff} \approx 1$, irrespective of the values of the parameters (see Fig.\ref{AHE-line}(b)). 
On the other hand, the magnitudes of
the peaks depend on two parameters: $\sigma^{AH}\sim(\Delta_R\tau)^2(\epsilon_{_F}\tau)^{-1}$.
 Figure \ref{AHE-cont} shows a contour plot of 
$ \tilde\sigma^{AH}=\sigma^{AH}(\epsilon_{_F}\tau)(\Delta_R\tau)^{-2}$ as a function of $\frac{q_x}{k_{_F}}$
and $(\Delta_{ex}\tau)^{1/2}( \epsilon_{_F}\tau)^{-1}$, {\it i.e.,} $(k{_F}l_{eff})^{-1}$.
 Note that for the nearly-linear darkest patch, in which the magnitude of $\tilde \sigma^{AH}$ is maximum, 
corresponds to $q_xl_{eff}\approx1$.

\begin{figure}
\includegraphics[width=\columnwidth]{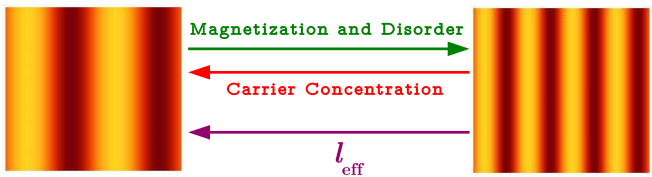}
\caption{(Color online) Schematic variation of the inhomogeneous electric field $E_x$, causing {\it resonance in AHC} : 
The periodicity  is illustrated by the color gradient. For increased magnetization and disorder and
decreased Fermi energy, {\it i.e.,}, for increased $l_{eff}$,
 the spatial variation of the electric field should be increased to obtain this resonance.}
\label{elec-field}
\end{figure}

These results suggest that if an electric field of the form $E_x=E_0 e^{iqx}$ is applied along
the x-axis in the plane of a two dimensional electron gas, the AHC becomes finite. As the maximum magnitude of  AHC occurs at
$q_x/k_F= (\Delta_{ex}\tau)^{1/2}(\epsilon_{_F}\tau)^{-1}$, the periodic variation of the applied field should be
properly tuned in accordance with the relevant parameters of the system to increase the transverse charge current. Clearly,
the nature of this periodicity depends on magnetization, disorder and carrier concentration
($i.e.$ on $\Delta_{ex},\tau$, and $\epsilon_{_F}$ respectively) and {\it not} on Rashba parameter $\alpha$.
This is illustrated in Fig. \ref{elec-field}. An increased variation of the electric field is necessary to produce maximum
AHC in a system with higher $l_{eff}$, {\i.e.}, with increasing magnetization and disorder, and decreasing electron density.

\section{summary}
To summarize, we have calculated the anomalous Hall conductivity for a two dimensional ferromagnetic Rashba system with
an electric field inhomogeneous in nature. It vanishes for a homogeneous electric field in the weak scattering regime
when both the subbands are occupied. We have shown that the anomalous Hall current can increase considerably due the variation
of the electric field and this comes from the bare bubble and ladder vertex correction and {\it{not}} from the skew scattering.
Dealing with a realistic parameter range, we have found that the length scale associated with the peaks in the conductivity is
proportional to the geometric mean of the two length scales arising from disorder and the exchange interaction. 

\begin{acknowledgments}
AS acknowledges the assistance provided by the scholars of theoretical physics department of IACS.
\end{acknowledgments}


\end{document}